\documentclass[10pt]{iopart}

\usepackage{graphicx}
\usepackage{amssymb}
\usepackage{dcolumn}
\usepackage{bm}
\usepackage{color}

\begin{document}
\title{Pressure-induced superconductivity in the giant Rashba system BiTeI}
\author{D.\ VanGennep,$^1$ A.\ Linscheid,$^1$ D.\ E.\ Jackson,$^1$ S.\ T.\ Weir,$^2$ Y.\ K.\ Vohra,$^3$ H.\ Berger,$^4$ G.\ R.\ Stewart,$^1$ R.\ G.\ Hennig,$^5$ P.\ J.\ Hirschfeld,$^1$ J.\ J.\ Hamlin$^1$}
\address{$^1$Department of Physics, University of Florida, Gainesville, FL 32611}
\address{$^2$Physics Division, Lawrence Livermore National Laboratory, Livermore, CA 94550, USA}
\address{$^3$Department of Physics, University of Alabama at Birmingham, Birmingham, AL, 35294}
\address{$^4$Institute of Physics, \'{E}cole Polytechnique F\'{e}d\'{e}rale de Lausanne, CH-1015 Lausanne, Switzerland}
\address{$^5$Department of Materials Science, University of Florida, Gainesville, FL 32611}
\ead{jhamlin@ufl.edu}

\date{\today}

\begin{abstract}
At ambient pressure, BiTeI is the first material found to exhibit a giant Rashba splitting of the bulk electronic bands.  At low pressures, BiTeI undergoes a transition from trivial insulator to topological insulator.  At still higher pressures, two structural transitions are known to occur.  We have carried out a series of electrical resistivity and AC magnetic susceptibility measurements on BiTeI at pressure up to $\sim 40$~GPa in an effort to characterize the properties of the high-pressure phases.  A previous calculation found that the high-pressure orthorhombic $P4/nmm$ structure BiTeI is a metal.  We find that this structure is superconducting with $T_c$ values as high as 6~K.  AC magnetic susceptibility measurements support the bulk nature of the superconductivity.  Using electronic structure and phonon calculations, we compute $T_c$ and find that our data is
consistent with phonon-mediated superconductivity.
\end{abstract}

\maketitle
\ioptwocol

\section{Introduction}
Over the past several years, the compounds BiTe$X$ ($X$ = Cl, Br, or I) have been the subject of a number of studies.  Interest in these compounds first surged when it was discovered via spin- and angle-resolved photoemission spectroscopy that BiTeI displays an enormous Rashba-like spin splitting of the bulk electronic bands~\cite{ishizaka_2011_1}.  Similar behavior was also observed in the Cl and Br analogues~\cite{Landolt_2013_1,Sakano_2013_1}. Although large Rashba splittings had previously been observed for surfaces~\cite{lashell_1996_1,koroteev_2004_1}, interfaces~\cite{nitta_1997_1}, and thin films~\cite{dil_2008_1,hirahara_2006_1,hirahara_2007_1}, BiTeI was the first material found to exhibit such large splittings in the bulk.  These materials may be useful for applications because they permit the creation and manipulation of spin polarized currents, and could allow the development of spintronic devices based on non-magnetic materials~\cite{marchenko_2012_1,takayama_2012_1}.

Interest in the BiTe$X$ family of compounds further increased following a first principles prediction by Bahramy \textit{et al.}~\cite{bahramy_2011_1} that BiTeI should undergo a band inversion and transition from trivial insulator to topological insulator under the application of a modest pressure of a few GPa.  The strong-spin orbit coupling and lack of inversion symmetry in this compound was predicted to lead to conspicuously different surface states on opposite sides of the material. Subsequently, a number of experimental efforts sought to find evidence for the proposed pressure-induced band inversion.  Infrared spectroscopy~\cite{xi_2013_1} and quantum oscillation data~\cite{vangennep_2014_1,Park2015} are consistent with the existence of a band inversion at 3-5 GPa.  However, given the limitations on the types of probes that can be applied at high pressure, no experimental effort has yet directly accessed the the surface states in the high pressure topologically non-trivial state.  On the other hand, Dirac surface states have been observed in BiTeCl, which appears to exist in the topologically non-trivial state at ambient pressure~\cite{Chen_2013_3}.

Some properties of BiTe$X$ compounds have also been probed at substantially higher pressures.  For example, a recent paper reported the results of electrical transport and Raman spectroscopic measurements on BiTeCl to 50 GPa~\cite{Ying_2016_1}. Changes in the Raman spectra near 5 and 35 GPa are suggestive of structural transitions at these pressures.  The electrical resistivity measurements show a dramatic increase in the resistivity upon increasing pressure above 5 GPa and a superconducting transition that appears at 10 GPa and reaches a maximum of $T_c \sim 8\,\mathrm{K}$ near 15-20 GPa.  This work proposed that BiTeCl-I (0-5 GPa) is a semiconductor, BiTeCl-II (5-35 GPa) an insulator, and BiTeCl-III a metal, with superconductivity appearing in both the insulating (II) and metallic (III) phases. While nominally an insulator, it is probable that phase-II of BiTeCl is actually a low carrier density metal, through \textit{e.g.,} site disorder, and is therefore capable of supporting a superconducting state.  In another work, the crystal structures of BiTeI were determined up to 30 GPa using high pressure x-ray diffraction~\cite{Chen_2013_2}.  It was found that BiTeI makes transitions from the ambient pressure BiTeI-I phase ($P3m1$) to BiTeI-II ($Pnma$) near 9 GPa, with an additional structural transition to BiTeI-III ($P4/nmm$) occurring near 19 GPa.

In the present work, we report electrical resistivity and magnetic susceptibility measurements on BiTeI to pressures as high as $\sim 40$ GPa.  These measurements show that superconductivity appears in the high pressure metallic BiTeI-III phase.  The size of the transition in the AC susceptibility is consistent with $100\%$ shielding, which rules out impurity phases as the source of the superconductivity.

To understand these results, density functional theory (DFT) calculations of electronic structure, electron-phonon interaction in the high pressure phase-III up to 40 GPa using Quantum Espresso and superconducting critical temperatures from the McMillan equation. Both the magnitude and pressure dependence of $T_c$ are qualitatively consistent with experiment, suggesting that the superconductivity in this high pressure phase is driven by the conventional electron-phonon interaction.

\section{Methods}
Single crystals of BiTeI were grown by the chemical vapor transport method. Small pieces of sample with dimensions of about  $70\,\mathrm{\mu m} \times 70\,\mathrm{\mu m} \times 10\,\mathrm{\mu m}$ were cut from a larger crystal for each of the measurements. The zero-field resistivity measurements as well as the AC magnetic susceptibility (ACS) measurements were carried out in a OmniDAC gas membrane-driven diamond anvil cell from Almax-EasyLab. The cell was placed inside a custom, continuous flow cryostat built by Oxford Instruments. Optical access to the cell was provided through windows at the bottom of the cryostat and an optical fiber entering through a feed-through at the top, allowing pressure to be measured at low temperature. The pressure was calibrated using the fluorescence of the $R_1$ peak of small ruby spheres placed next to the sample~\cite{piermarini_1975_1}. The high-field resistivity measurements were performed in a Quantum Design Physical Property Measurement System (PPMS) using an Almax-EasyLab ChicagoDAC.

For the resistivity measurements, one of the diamonds used was a designer diamond anvil containing eight symmetrically arranged, deposited tungsten microprobes encapsulated in high-quality homoepitaxial diamond~\cite{weir_2000_1}. This diamond had a culet diameter of $\sim 180\,\mathrm{\mu m}$, and the opposing anvil had a culet diameter of $\sim 500\,\mathrm{\mu m}$. Resistance was measured in the crystalline $ab$-plane by either the Quantum Design PPMS or a Lakeshore Model 370 AC resistance bridge using the four-probe Van der Pauw method with currents of $\leq 1\,\mathrm{mA}$. In the high-field measurements, the field was applied along the $c$-axis. Gaskets were preindented from $250\,\mathrm{\mu m}$ to $\sim 30\,\mathrm{\mu m}$ thickness and were made of 316 stainless steel for the resistance measurements, and of a BeCu alloy for the AC susceptibility measurements. Quasihydrostic soft, solid steatite was used as the pressure-transmitting medium for the resistance measurements, while a 1:1 mixture of n-pentane:isoamyl alcohol~\cite{Butch_2009_1} was used for the AC susceptibility measurements.

For the AC susceptibility measurements, the superconducting transitions were determined inductively using a balanced primary/secondary coil system~\cite{deemyad_2001_1} located immediately outside the metal gasket and connected to a Stanford Research SR554 transformer pre-amplifier and a Stanford Research SR830 digital lock-in amplifier. The sample was subject to an AC magnetic field of $\sim 3\,\mathrm{Oe}$ RMS applied along the $c$-axis with a frequency of $\sim 1\,\mathrm{kHz}$.

\begin{figure}[t]
  \centering
  \includegraphics[width=\columnwidth]{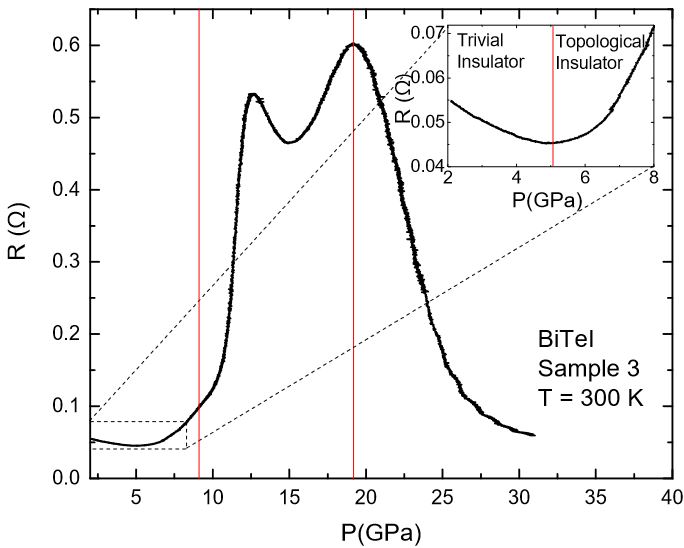}
  \caption{(Color online) Electrical resistance {\it vs.} pressure for BiTeI at room temperature.  The red vertical lines indicate the locations of the previously reported structural transitions~\cite{Chen_2013_2}.  Inset highlights a resistivity minimum in the vicinity of the predicted topological transition.}
  \label{fig:R_P}
\end{figure}

We have performed a calculation of the electronic structure using density-functional theory employing a plane-wave basis, the PBEsol exchange-correlation functional~\cite{Perdew2008}, and ultrasoft pseudo potentials~\cite{Garrity2014} as implemented in the quantum ESPRESSO package~\cite{QE-2009}.  The phonon dispersion and electron-phonon coupling were calculated using density-functional perturbation theory \cite{Baroni1987}. We use a plane-wave and charge cutoff of 42 and 168~Ry, respectively, which ensures a total energy convergence of 1 mRy/atom. For the electronic structure and phonon calculation, we sample the Brillouin zone with a regular $8\times8\times4$ $k$-points mesh with a Marzari-Vanderbilt smearing~\cite{Marzari1999} of 0.01~Ry. This sampling yields total energy convergence of 1 mRy/atom and for this smearing and matches the converged total energy of a higher sampling with lesser smearing. The structure was relaxed with a force and energy convergence of $5\times10^{-5}$~Ry/a$_\mathrm{Bohr}$ and $10^{-6}$ Ry respectively. For the phonon spectrum, we perform calculations for the structure at 25~GPa up to $8\times8\times4$  $q$-point meshes and determined that a $6\times6\times3$ mesh is sufficiently accurate. We calculate the phonon spectra at pressures of 25, 30, 35, and 40~GPa using a $6\times6\times3$ sampling mesh. For the electron-phonon calculation, we have increased the Brillouin sampling mesh to $16\times16\times10$ for the electronic wave function.  The resulting electron phonon coupling is then averaged on the Fermi surface using a metropolis algorithm with $60000$ random $k$ points.

The resulting Eliashberg function was used to compute the logarithmic frequency $\omega_{\rm{ln}}$ and the effective electron-phonon coupling parameter $\lambda_{\rm{ep}}$~\cite{CarbottePropertiesOfBosonExchangeSC1990}. $T_c$ was calculated from the McMillan equation using these parameters together with a reasonable range for the effective Coulomb interaction.

\section{Experimental results}
Figure~\ref{fig:R_P} shows a typical measurement of the electrical resistivity \textit{vs}.\ pressure at room temperature.  The data were collected using a recently completed system that automatically calculates pressure from the fluorescence spectrum of the ruby in real time.  This makes it possible to collect a great number of data points while the pressure is slowly swept upwards.  This system will be detailed in a future publication.

The inset of Figure~\ref{fig:R_P} shows that at low pressures, the resistivity initially decreases, before passing through a minimum at about 5 GPa.  This is near the pressure where BiTeI is thought to undergo a band inversion accompanied by a transition from trivial to topological insulator.  At the critical pressure for the band inversion, $P_c$, a near linear dispersion is expected, which would lead to a minimum in the effective mass at $P_c$.  All other things being equal, the minimum in effective mass and closing of the band gap should result in a conductivity maximum (resistivity minimum) consistent with our data.

At higher pressures, other features appear in the resistivity that are likely connected with structural transitions.  The locations of the known structural transitions are indicated by vertical red lines.  Near the transition from BiTeI-I (space group: $P3m1$) to BiTeI-II (space group: $Pnma$), the resistivity abruptly begins to increase before then decreasing in the BiTeI-III structure (space group: $P4/nmm$).  These trends are consistent with previous electronic structure calculations~\cite{Chen_2013_2} that predicted that (1) BiTeI-II is a semiconductor with a larger band gap than BiTeI-I, and (2) BiTeI-III is a metal.

\begin{figure}
  \centering
  \includegraphics[width=\columnwidth]{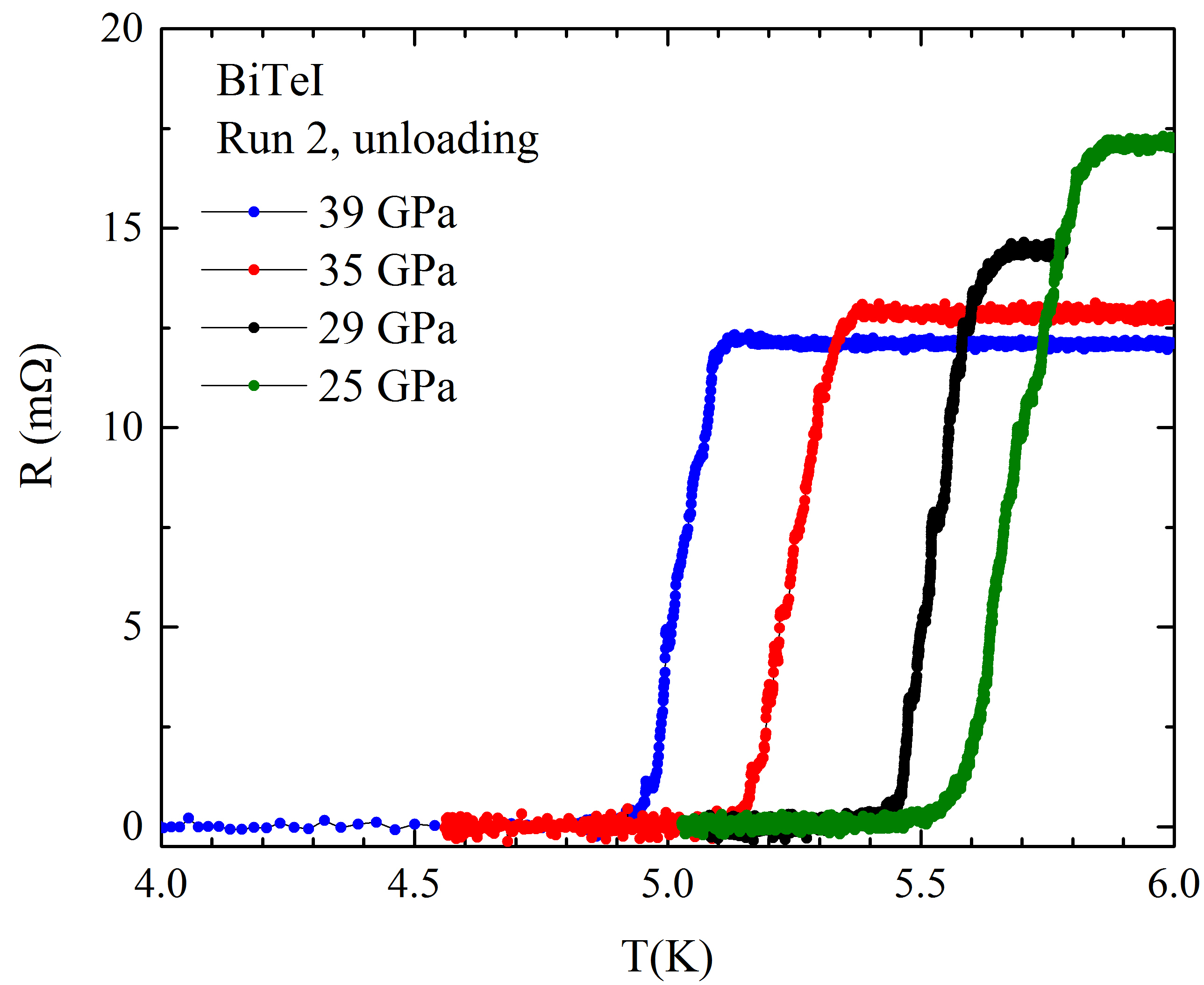}
  \caption{(Color online) Resistivity \textit{vs}.\ temperature for BiTeI at several different values of the applied pressure showing complete superconducting transitions.  The data shown were collected while decreasing (unloading) from high pressure.}
  \label{fig:R_T}
\end{figure}

The metallic nature of BiTeI-III is further supported by the occurrence of superconductivity in this phase.  Figure~\ref{fig:R_T} shows electrical resistivity \textit{vs}.\ temperature data from ``Run 2.''  The onset of the superconducting transitions reaches as high as about 5.8 K.  The transitions are relatively sharp with a width of $\sim 3\%$ of $T_c$, and the resistivity appears to drop to zero at low temperature.  For this data, the slope of $T_c$ \textit{vs}.\ pressure is $dT_c/dP \sim -0.05\, \mathrm{K/GPa}$.

Despite the complete transition to zero resistance at low temperatures, it is important to consider the possibility that the superconductivity could derive from an impurity phase that percolates through the sample.  This is particularly important to consider given that elemental bismuth, tellurium, and iodine all become superconductors under pressure~\cite{Ilina1972,Akahama1992,Shimizu1994}.  Though the $T_c$ of iodine never exceeds 1.2 K, bismuth has a $T_c$ of about $6\, \mathrm{K}$ at 20~GPa and tellurium has a $T_c$ of about $7.5\, \mathrm{K}$ at 35~GPa.  In order to determine whether the observed superconductivity could be attributed to impurity phases, we carried out AC magnetic susceptibility measurements.

\begin{figure}[t]
  \centering
  \includegraphics[width=\columnwidth]{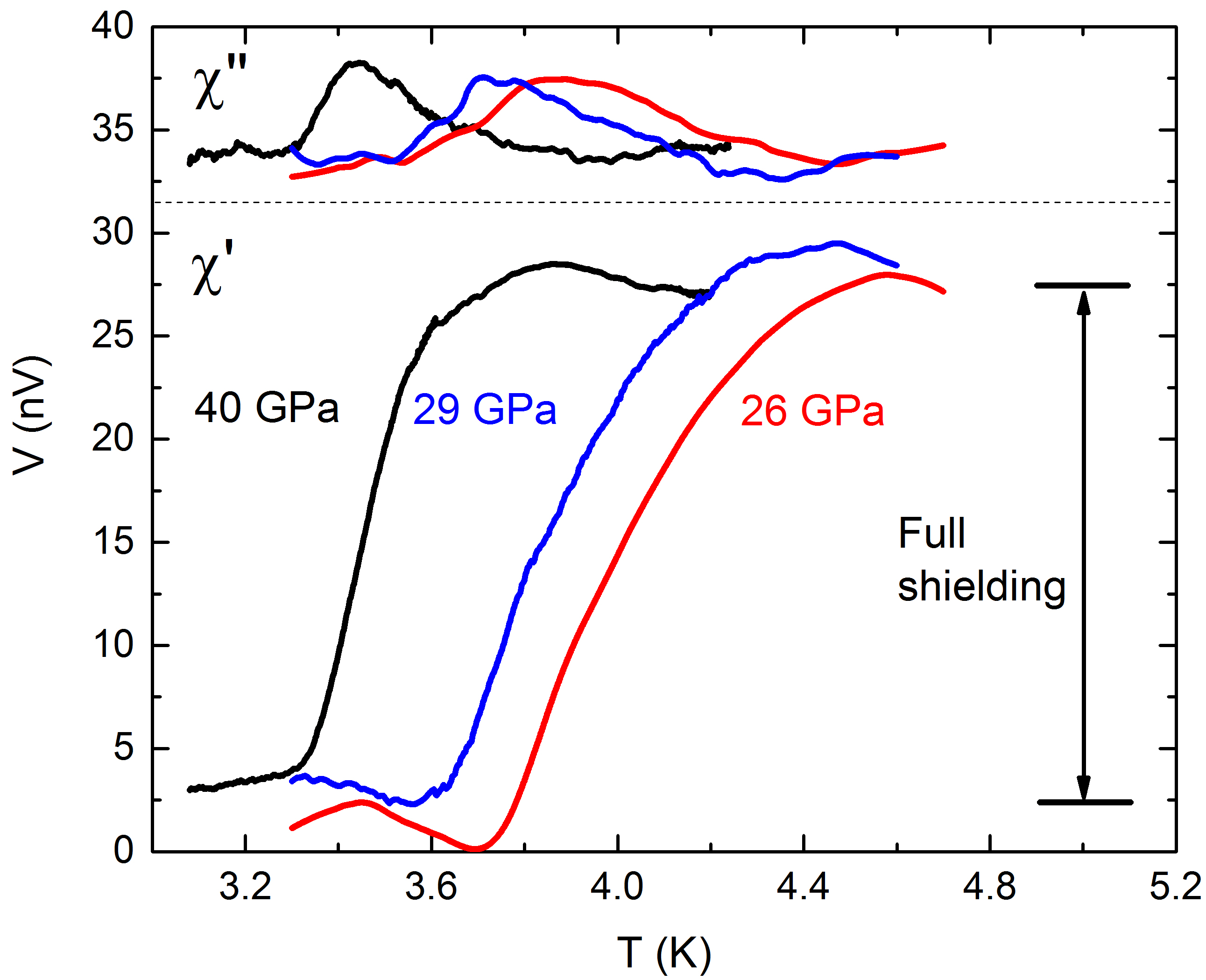}
  \caption{(Color online) Real ($\chi ^{\prime}$) and imaginary ($\chi ^{\prime \prime}$) parts of the AC magnetic susceptibility {\it vs.} temperature for BiTeI at high pressure.  The interval labeled ``Full shielding'' at the right of the figure indicates the expected size of the transition for $100 \%$ expulsion of flux from the sample.  The results are thus consistent with bulk superconductivity in BiTeI at high pressures.}
  \label{fig:chi}
\end{figure}

Figure~\ref{fig:chi} presents both the real ($\chi ^{\prime}$) and imaginary ($\chi ^{\prime \prime}$) parts of the AC magnetic susceptibility \textit{vs}.\ temperature for 26, 29, and 40~GPa.  The data have been plotted in nV, indicating the induced voltage in the pickup coil.  The clear drops in $\chi ^{\prime}$ accompanied by peaks in $\chi ^{\prime \prime}$ are typical of a superconductor.  The slope of $dT_c/dP \sim -0.04\, \mathrm{K/GPa}$, which is nearly the same as the slope observed in the resistivity measurements.  The interval labeled ``Full shielding'' in Figure~\ref{fig:chi} shows the expected size of the transition assuming bulk superconductivity and is estimated using the geometry of the coil system, frequency and magnitude of the applied AC field, and the geometry of the sample (including the demagnetization factor).  The data clearly suggest that the superconductivity derives from bulk BiTeI, rather than any impurity phase.  We also note that the slopes of $dT_c/dP$ for elemental bismuth and tellurium are roughly four and nine times larger, respectively, than we observe in the pressure range of interest, which provides further evidence against the possibility of impurity superconductivity.

\begin{figure}
  \centering
  \includegraphics[width=\columnwidth]{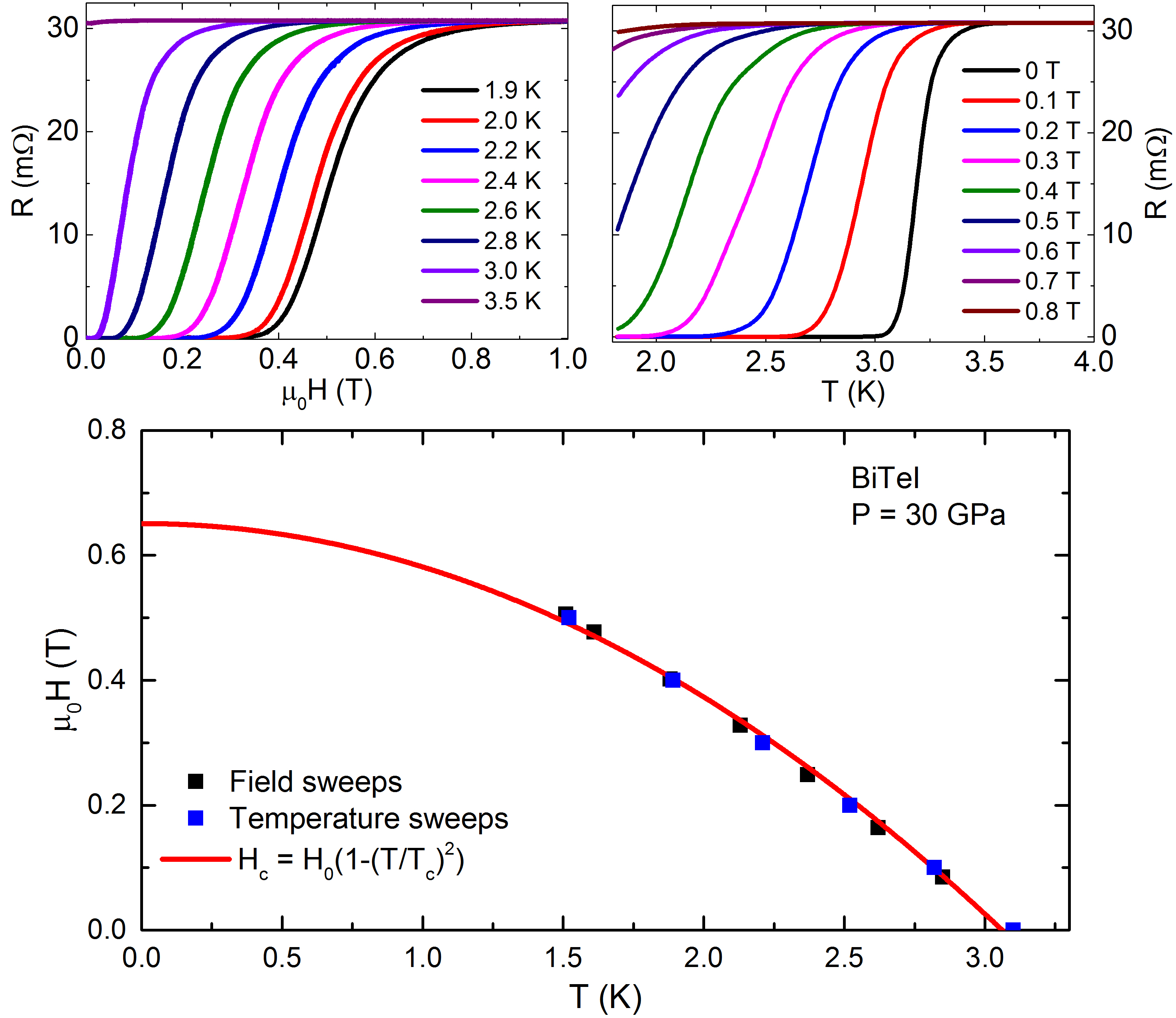}
  \caption{Influence of applied magnetic fields on the superconducting transition in BiTeI at $P \sim 30\,\mathrm{GPa}$. \textit{upper left:} Resistivity \textit{vs}.\ applied field at several temperatures. \textit{upper right:} Resistivity \textit{vs}.\ temperature at several applied magnetic fields. \textit{lower panel:} Field \textit{vs}.\ transition temperature.  Extrapolation to zero temperature yields a critical field of $\sim 0.65\,\mathrm{tesla}$ for $T_c = 3.1\,\mathrm{K}$.}
  \label{fig:Hc}
\end{figure}

In order to further characterize the superconducting state in BiTeI, we carried out measurements in applied magnetic fields at a pressure of $\sim 30\,\mathrm{GPa}$.  In this particular experiment $T_c \sim 3\,\mathrm{K}$, which is somewhat lower than in our other experiments at similar pressures.  The fields were applied along the crystallographic $c$-axis.  Fields of less than 1 tesla are sufficient to completely suppress the superconducting state.  Figure~\ref{fig:Hc} presents a summary of the high field measurements.  The upper left and right panels show field sweeps at constant temperature and temperature sweeps at constant field, respectively.  The transition remains rather sharp as it is suppressed.

The lower panel of Figure~\ref{fig:Hc} shows field \textit{vs}.\ transition temperature with data points taken both from field sweeps and temperature sweeps.  Extrapolation of the critical field curve to zero temperature using the relation $H_c = H_{c,0}[1-(T/T_c)^2]$ yields $H_c \sim 0.65\,\mathrm{tesla}$.  Alternatively, a WHH analysis~\cite{Werthamer_1966_1} $H_{c2}(0)=-0.7\,T_c\,(dH_{c2}/dT)|_{T=T_c}$ yields a zero temperature critical field of $0.56\,\mathrm{tesla}$. Both of these values are well below the weak coupling BCS paramagnetic limit $\mu_0 H_p^\mathrm{BCS} = 1.84\,T_c$, where $\mu_0 H_p^\mathrm{BCS}$ is in units of tesla and $T_c$ is in units of kelvin.

\begin{figure}
  \centering
  \includegraphics[width=\columnwidth]{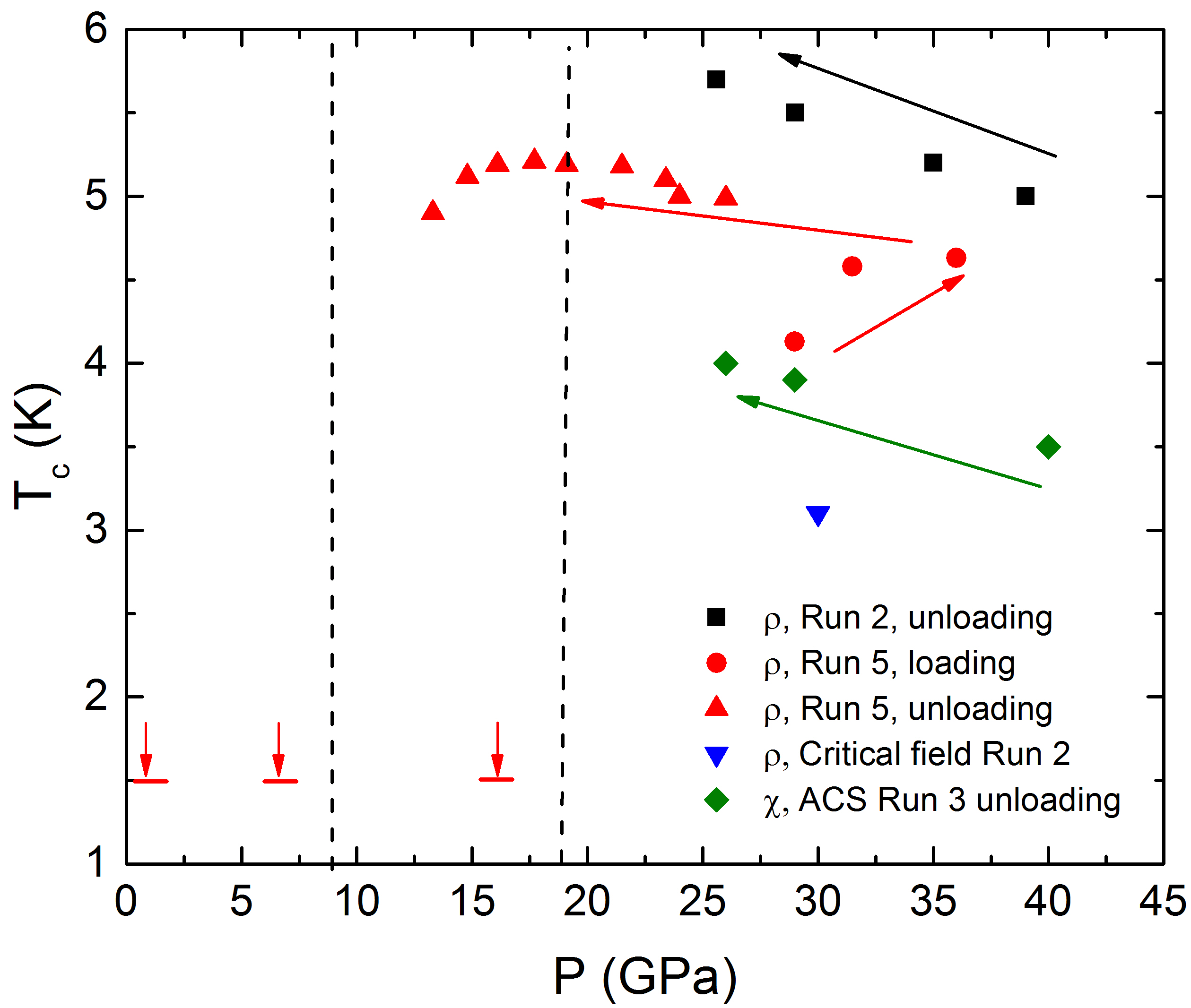}
  \caption{$T_c$ \textit{vs}.\ pressure phase diagram for BiTeI.  The vertical dashed lines indicate the previously reported locations of the structural phase boundaries~\cite{Chen_2013_2} (determined at room temperature).  Arrows indicate the order of measurement.  The vertical arrows at low pressure indicate that no superconductivity was observed down to 1.5 K at those pressures during loading.}
  \label{fig:phase}
\end{figure}

Figure~\ref{fig:phase} presents a phase diagram of $T_c$ \textit{vs}.\ pressure that summarizes the data from several experiments.  The dashed vertical lines indicate the locations of the structural phase transitions at room temperature.  The black, red, and blue data points represent transitions measured using electrical resistance measurements, while the green data points were collected via ac magnetic susceptibility.  Although the exact values of $T_c$ vary somewhat from one measurement to the next, the overall trends are the same, and indeed the slopes, $dT_c/dP$, are nearly identical.  The scatter in the data might be due to the pressure conditions, which are only quasi-hydrostatic, and thus may vary somewhat from one experiment to the next.

During increasing pressure, we have not observed superconductivity at pressures below $\sim 28\,\mathrm{GPa}$ (see red data points, ``Run 5'').  When increasing pressure to $\sim 16\,\mathrm{GPa}$ at room temperature, we see no trace of superconductivity down to $\sim 1.5\,\mathrm{K}$. When increasing pressure, we initially observe an increase in $T_c$ with pressure.  It is possible that this effect is due to a sluggish structural transition on loading.  During unloading from higher pressures, we always observe an increase in $T_c$ with decreasing pressure.  When releasing the pressure at low temperature ($T \lesssim 10\,\mathrm{K}$), the superconductivity persists to lower pressures than it can be observed during pressure loading.  Again, this is probably related to the sluggish or broad nature of the II-III (or III-II) structural transition.  During unloading at low temperature (Run 5), we find that superconductivity persists all the way down to about 13 GPa, presumably indicating that the system remains in a metastable state of phase III.

\begin{figure}
\begin{center}
\includegraphics[width=1\columnwidth]{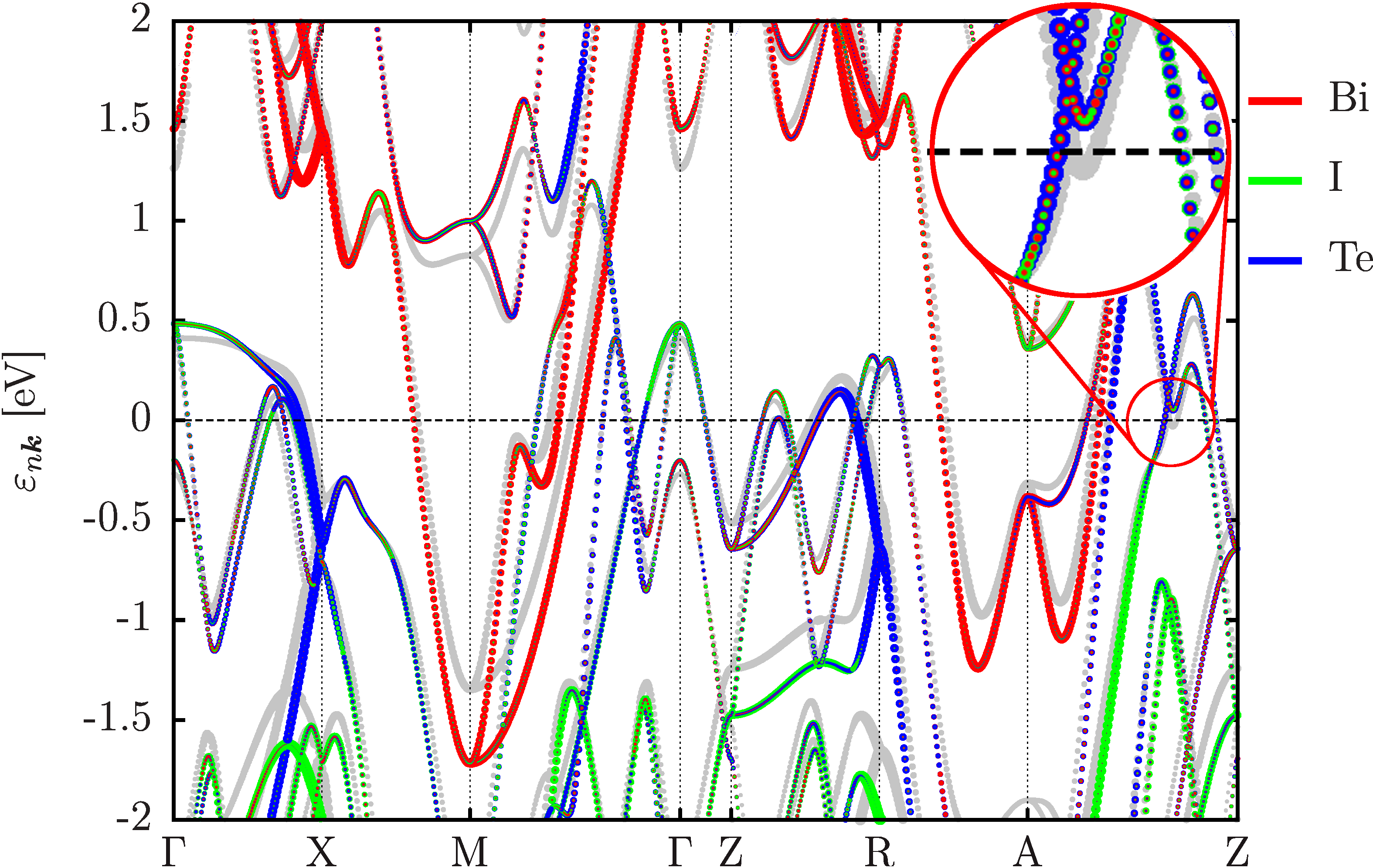}\\
\par\end{center}%

\caption{Band structure at $40$ GPa with
a projection on atomic states together with the one $25$ GPa as a light gray in the background. The color of the dots indicates the
type of atom while their size indicates the overlap with the respective atomic states.\label{fig:DFTBands}}

\end{figure}

\section{Computational results}
Figure~\ref{fig:DFTBands} shows that the electronic band structure of BiTeI at pressures of 25 and 40 GPa is dominated near the Fermi level by Bi states, and that several strongly hybridized states cross the Fermi level. The band structures at $25$ and $40$ GPa are fairly similar. One notable feature is an additional electron-like band crossing the Fermi level, visible between the A and Z point in Fig.~\ref{fig:DFTBands}, as we pressurize the system from $25$ to $40$ GPa.

\begin{figure}
\begin{centering}
\includegraphics[width=1\columnwidth]{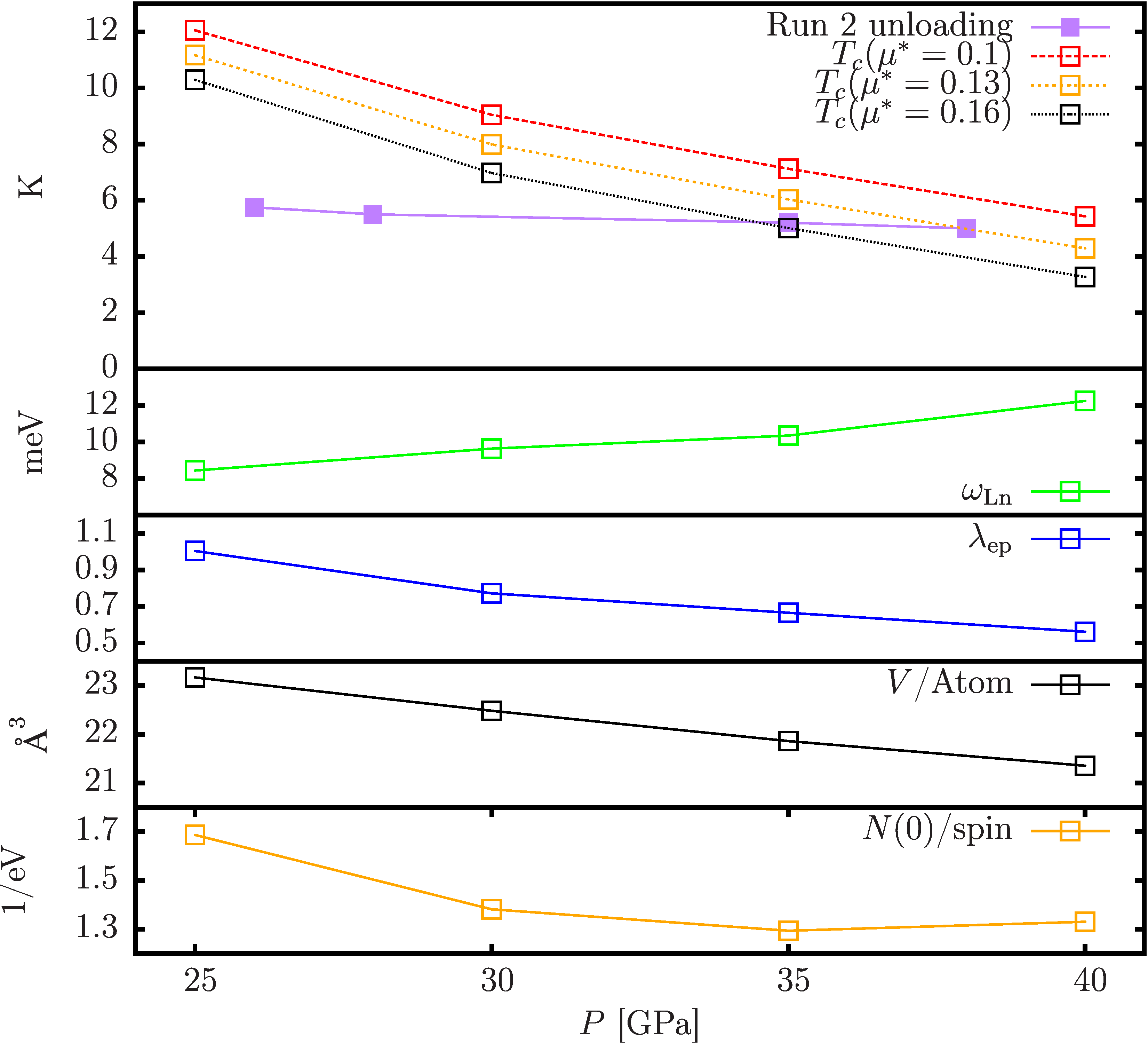}
\par\end{centering}

\caption{Calculated $T_{c}$ for different screened Coulomb interaction parameters together with experimental data, logarithmic frequency and electron-phonon coupling unit cell volume and DOS at the Fermi level as a function of pressure.\label{fig:CalcTcWlnLambda}}

\end{figure}

We show the calculated $T_c$ for different $\mu^\ast$ together with $T_c$ from one of the experimental runs, the logarithmically averaged phonon frequency $\omega_{{\rm ln}}$, the effective electron-phonon coupling values $\lambda_{{\rm ep}}$, the volume $V$ of the unit cell, and the density of states at the Fermi level $N(0)$ per spin as a function of pressure in Fig.~\ref{fig:CalcTcWlnLambda}.
The coupling increases as the pressure is reduced while the coupling
frequency decreases, indicating a softening of the relevant coupling
modes. Much of the increase of the coupling from $30$ to $25$ GPa, and thus the increasing
$T_{c}$, has to be attributed to the increasing DOS with lower pressure. From the band structure in Fig.~\ref{fig:DFTBands}, we attribute this to the additional predominantly Te-like Fermi surface that appears upon lowering the pressure.
Especially at high pressures we see a good agreement between experiment and theoretical calculations. Also, we reproduce the observed trend to higher $T_{c}$ as we lower the pressure and approach the structural transition. Since we find a rather large coupling constant, $T_c$ is not very sensitive to the Coulomb repulsion parameter.
The agreement between experiment and theory for the magnitude of $T_c$ and
 the observed increase in $T_c$ at lower pressure when BiTeI approaches the structural phase transition, demonstrates that the superconductivity BiTeI under pressure occurs by a conventional phonon mediated mechanism.

\section{Discussion}
It is interesting to compare the phase diagram of BiTeI to that previously reported for BiTeCl~\cite{Ying_2016_1}.  Both compounds exhibit a very similar evolution under increasing pressure from a semiconducting structure, to an intermediate semiconducting structure with higher resistivity, and finally to a metallic structure.  In the case of BiTeI, the evidence seems to suggests that the intermediate high resistivity phase (BiTeI-II) is not superconducting.  Although superconductivity is sometimes observed in the pressure range assigned to BiTeI-II, this only occurs after unloading from pressures high enough to enter the BiTeI-III structure.  Thus, the superconductivity observed in the BiTeI-II pressure range can likely be attributed to portions of the sample remaining in the phase III high pressure structure.

In contrast, in the case of BiTeCl, recent work shows superconductivity appearing in the intermediate, insulating, phase II structure~\cite{Ying_2016_1}.  It is possible that, in the chloride, phase-II is superconducting, while in the iodide, phase II is non-superconducting.  However, it is worth pointing out that the critical pressures for the structural transitions are based primarily on room temperature data and that the phase boundaries could move as temperature is lowered.  Low temperature x-ray diffraction measurements on BiTeI and BiTeCl would be helpful in clarifying whether phase II is superconducting in either compound.

\section{Conclusion}
We have carried out a series of high pressure magnetic susceptibility and electrical resistivity measurements on single crystals of BiTeI.  These measurements show that the metallic high-pressure phase-III of BiTeI (space group: $P4/nmm$) is a bulk superconductor. First principles electronic structure and phonon calculations are able to reproduce the approximate value of $T_c$ and trend of decreasing $T_c$ with increasing pressure, indicating that the superconductivity arises due to a conventional electron-phonon mechanism.\\

\noindent \textit{Note added:} In the final stages of writing, a paper reporting a similar observation of superconductivity in BiTeI at high-pressure appeared as arXiv:1610.05364.

\section{Acknowledgments}
Development of in situ pressure tuning equipment partially supported The National High Magnetic Field Laboratory User Collaboration Grants Program.  The National High Magnetic Field Laboratory is supported by National Science Foundation Cooperative Agreement No. DMR-1157490 and the State of Florida.  Measurements supported by DMR-1453752, theoretical work supported by  U.S. Department of Energy DE-FG02-05ER46236, designer diamond anvils supported by DOE-NNSA Grant No.\ DE-NA0002928 and under the auspices of the U.S. Department of Energy by Lawrence Livermore National Laboratory under Contract DE-AC52-07NA27344, GS supported by the US DOE, Office of Basic Energy Sciences, contract no.\ DE-FG02-86ER45268.

\section{References}
\providecommand{\newblock}{}


\begin{thebibliography}{10}
\expandafter\ifx\csname url\endcsname\relax
  \def\url#1{{\tt #1}}\fi
\expandafter\ifx\csname urlprefix\endcsname\relax\def\urlprefix{URL }\fi
\providecommand{\eprint}[2][]{\url{#2}}

\bibitem{ishizaka_2011_1}
Ishizaka K, Bahramy M~S, Murakawa H, Sakano M, Shimojima T, Sonobe T, Koizumi
  K, Shin S, Miyahara H, Kimura A, Miyamoto K, Okuda T, Namatame H, Taniguchi
  M, Arita R, Nagaosa N, Kobayashi K, Murakami Y, Kumai R, Kaneko Y, Onose Y
  and Tokura Y 2011 {\em Nature Materials\/} {\bf 10} 521

\bibitem{Landolt_2013_1}
Landolt G, Eremeev S~V, Tereshchenko O~E, Muff S, Slomski B, Kokh K~A,
  Kobayashi M, Schmitt T, Strocov V~N, Osterwalder J, Chulkov E~V and Dil J~H
  2013 {\em New Journal of Physics\/} {\bf 15} 085022

\bibitem{Sakano_2013_1}
Sakano M, Bahramy M~S, Katayama A, Shimojima T, Murakawa H, Kaneko Y, Malaeb W,
  Shin S, Ono K, Kumigashira H, Arita R, Nagaosa N, Hwang H~Y, Tokura Y and
  Ishizaka K 2013 {\em Physical Review Letters\/} {\bf 110} 1--5

\bibitem{lashell_1996_1}
LaShell S, McDougall B~A and Jensen E 1996 {\em Phys. Rev. Lett.\/} {\bf 77}
  3419

\bibitem{koroteev_2004_1}
Koroteev Y~M, Bihlmayer G, Gayone J~E, Chulkov E~V, Bl{\"u}gel S, Echenique P~M
  and Hofmann P 2004 {\em Phys. Rev. Lett.\/} {\bf 93} 046403

\bibitem{nitta_1997_1}
Nitta J, Akazaki T, Takayanagi H and Enoki T 1997 {\em Phys. Rev. Lett.\/} {\bf
  78} 1335

\bibitem{dil_2008_1}
Dil J~H, Maier F, Lobo-Checa J, Patthey L, Bihlmayer G and Osterwalder J 2008
  {\em Phys. Rev. Lett.\/} {\bf 101} 266802

\bibitem{hirahara_2006_1}
Hirahara T, Nagao T, Matsuda I, Bihlmayer G, Chulkov E~V, Koroteev Y~M,
  Echenique P~M, Saito M and Hasegawa S 2006 {\em Phys. Rev. Lett.\/} {\bf 97}
  146803

\bibitem{hirahara_2007_1}
Hirahara T, Nagao T, Matsuda I, Bihlmayer G, Chulkov E~V, Koroteev Y~M and
  Hasegawa S 2007 {\em Phys. Rev. B.\/} {\bf 75} 035422

\bibitem{marchenko_2012_1}
Marchenko D, Varykhalov A, Sholz M~R, Bihlmayer G, Rashba E~I, Rybkin A, Shikin
  A~M and Rader O 2012 {\em Nature Communications\/} {\bf 3} 1232

\bibitem{takayama_2012_1}
Takayama A, Sato T, Souma S, Oguchi T and Takahashi T 2012 {\em Nano Letters\/}
  {\bf 12} 1776

\bibitem{bahramy_2011_1}
Bahramy M~S, Yang B~J, Arita R and Nagaosa N 2011 {\em Nature Communications\/}
  {\bf 342} 679

\bibitem{xi_2013_1}
Xi X, Ma C, Liu Z, Chen Z, Ku W, Berger H, Martin C, Tanner D~B and Carr G~L
  2013 {\em Phys. Rev. Lett.\/} {\bf 111} 155701

\bibitem{vangennep_2014_1}
VanGennep D, Maiti S, Graf D, Tozer S~W, Martin C, Berger H, Maslov D~L and
  Hamlin J~J 2014 {\em J. Phys: Condens. Matter\/} {\bf 26} 342202

\bibitem{Park2015}
Park J, Jin K~H, Jo Y~J, Choi E~S, Kang W, Kampert E, Rhyee J~S, Jhi S~H and
  Kim J~S 2015 {\em Scientific Reports\/} {\bf 5} 15973

\bibitem{Chen_2013_3}
Chen Y~L, Kanou M, Liu Z~K, Zhang H~J, Sobota J~A, Leuenberger D, Mo S~K, Zhou
  B, Yang S~L, Kirchmann P~S, Lu D~H, Moore R~G, Hussain Z, Shen Z~X, Qi X~L
  and Sasagawa T 2013 {\em Nature Physics\/} {\bf 9} 704--708

\bibitem{Ying_2016_1}
Ying J~j, Struzhkin V~V, Cao Z~y, Goncharov A~F, Mao H~k, Chen F, Chen X~h,
  Gavriliuk A~G and Chen X~j 2016 {\em Physical Review B - Rapid
  Communications\/} {\bf 93} 1--6

\bibitem{Chen_2013_2}
Chen Y, Xi X, Yim W~L, Peng F, Wang Y, Wang H, Ma Y, Liu G, Sun C, Ma C, Chen Z
  and Berger H 2013 {\em Journal of Physical Chemistry C\/} {\bf 117}
  25677--25683

\bibitem{piermarini_1975_1}
Piermarini G~J, Block S, Barnett J~D and Forman R~A 1975 {\em J. Appl. Phys.\/}
  {\bf 46} 2774

\bibitem{weir_2000_1}
Weir S~T, Akella J, Aracne-Ruddle C, Vohra Y~K and Catledge S~A 2000 {\em Appl.
  Phys. Lett.\/} {\bf 77} 3400

\bibitem{Butch_2009_1}
Butch N~P, Jeffries J~R, Zocco D~A and Maple M~B 2009 {\em High Pressure
  Research\/} {\bf 29} 335--343

\bibitem{deemyad_2001_1}
Deemyad S, Schilling J~S, Jorgensen J~D,  and Hinks D~G 2001 {\em Physica C\/}
  {\bf 361} 227

\bibitem{Perdew2008}
Perdew J~P, Ruzsinszky A, Csonka G~I, Vydrov O~A, Scuseria G~E, Constantin L~A,
  Zhou X and Burke K 2008 {\em Physical Review Letters\/} {\bf 100} 136406

\bibitem{Garrity2014}
Garrity K~F, Bennett J~W, Rabe K~M and Vanderbilt D 2014 {\em Computational
  Materials Science\/} {\bf 81} 446--452

\bibitem{QE-2009}
Giannozzi P, Baroni S, Bonini N, Calandra M, Car R, Cavazzoni C, Ceresoli D,
  Chiarotti G~L, Cococcioni M, Dabo I, {Dal Corso} A, de~Gironcoli S, Fabris S,
  Fratesi G, Gebauer R, Gerstmann U, Gougoussis C, Kokalj A, Lazzeri M,
  Martin-Samos L, Marzari N, Mauri F, Mazzarello R, Paolini S, Pasquarello A,
  Paulatto L, Sbraccia C, Scandolo S, Sclauzero G, Seitsonen A~P, Smogunov A,
  Umari P and Wentzcovitch R~M 2009 {\em Journal of Physics: Condensed
  Matter\/} {\bf 21} 395502

\bibitem{Baroni1987}
Baroni S, Giannozzi P and Testa A 1987 {\em Physical Review Letters\/} {\bf 58}
  1861--1864

\bibitem{Marzari1999}
Marzari N, Vanderbilt D, {De Vita} A and Payne M~C 1999 {\em Physical Review
  Letters\/} {\bf 82} 3296--3299

\bibitem{CarbottePropertiesOfBosonExchangeSC1990}
Carbotte J~P 1990 {\em Rev. Mod. Phys.\/} {\bf 62} 1027--1157

\bibitem{Ilina1972}
Il'ina M~A, Itskevich E~S and Dizhur E~M 1972 {\em Soviet Physics JETP\/} {\bf
  34} 1263

\bibitem{Akahama1992}
Akahama Y, Kobayashi M and Kawamura H 1992 {\em Solid state communications\/}
  {\bf 84} 803--806

\bibitem{Shimizu1994}
Shimizu K, Yamauchi T, Tamitani N, Takeshita N, Ishizuka M, Amaya K and Endo S
  1994 {\em Journal of Superconductivity\/} {\bf 7} 921--924

\bibitem{Werthamer_1966_1}
Werthamer N~R, Helfand E and Hohenberg P~C 1966 {\em Physical Review\/} {\bf
  147} 295--302

\end{thebibliography}
\end{document}